\documentclass[]{jfm}

\usepackage{graphicx}
\usepackage{newtxtext}
\usepackage{newtxmath}
\usepackage{natbib}
\usepackage{hyperref}
\usepackage{comment}

\usepackage{caption}

\hypersetup{
    colorlinks = true,
    urlcolor   = blue,
    citecolor  = black,
}

\newcommand{\RomanNumeralCaps}[1]

\newcommand{\bk}{\boldsymbol{k}}

\newcommand{\eez}{\boldsymbol{e}_z}
\newcommand{\bu}{\boldsymbol{u}}
\newcommand{\ok}{\omega_{\bk}}
\newcommand{\pt}{\partial_t}
\newcommand{\px}{\partial_x}
\newcommand{\pz}{\partial_z}
\newcommand{\diff}{\mathrm{d}}

\newcommand{\hpsi}{\hat{\psi}_{\bk}}
\newcommand{\hbu}{\hat{\bu}_{\bk}}
\newcommand{\hb}{\hat{b}_{\bk}}
\newcommand{\hfpsi}{\hat{f}_{\psi,\bk}}
\newcommand{\hfb}{\hat{f}_{b,\bk}}

\newcommand{\kf}{k_{\rm f}}
\newcommand{\kfmin}{k_{\rm f,min}}
\newcommand{\kfmax}{k_{\rm f,max}}
\newcommand{\kmax}{k_{\rm max}}
\newcommand{\kb}{k_{\rm b}}
\newcommand{\ko}{k_{\rm O}}
\newcommand{\keta}{k_{\rm \eta}}
\newcommand{\kc}{k_{\rm c}}
\newcommand{\kd}{k_{\rm d}}

\newcommand{\ekin}{e_{\rm kin}}
\newcommand{\epot}{e_{\rm pot}}
\newcommand{\Tkin}{\mathcal{T}_{\rm kin}}
\newcommand{\Tpot}{\mathcal{T}_{\rm pot}}

\newcommand{\oemp}{\omega_{\rm emp}}

\newcommand{\mean}[1]{\left\langle #1 \right\rangle}

\title{2D Internal Gravity Wave Turbulence}

\author{V. Labarre\aff{1,3} \corresp{\email{vincent.labarre@polytechnique.edu}} \and M. Shavit\aff{2} \corresp{\email{ms14479@nyu.edu}}}

\affiliation{\aff{1}Université Côte d’Azur, Observatoire de la Côte d’Azur, CNRS, Laboratoire Lagrange, Boulevard de l’Observatoire CS 34229–F 06304 Nice Cedex 4, France \aff{2}Courant Institute of Mathematical Sciences, New York University, NY 10012, USA \aff{3}LadHyX, CNRS, \'{E}cole polytechnique, Institut polytechnique de Paris, 91120, Palaiseau, France}

\begin{document}
\maketitle

\begin{abstract}
	Using weak wave turbulence theory analysis, we distinguish three main regimes for 2D stratified fluids in the dimensionless parameter space defined by the Froude number and the Reynolds number: discrete wave turbulence, weak wave turbulence, and strong nonlinear interaction. These regimes are investigated using direct numerical simulations (DNS) of the 2D Boussinesq equations with shear modes removed. In the weak wave turbulence regime, excluding slow frequencies, we observe a spectrum that aligns with recent predictions from kinetic theory. This finding represents the first DNS-based confirmation of wave turbulence theory for internal gravity waves. At strong stratification, in both the weak and strong interaction regimes, we observe the formation of layers accompanied by spectral peaks at low discrete frequencies. We explain this layering through an inverse kinetic energy cascade and the discreteness of wave-wave interactions at large scales. This analysis allows us to predict the layer thickness and typical flow velocity in terms of the control parameters.
\end{abstract}


\section{Introduction}

Internal gravity waves propagating within stably stratified fluids are ubiquitous in geophysical and astrophysical systems. Through the transport of mass, momentum, and energy, they play a crucial role in shaping oceanic and atmospheric circulations \citep{andrews1987middle,buhler2014waves,vallis2017atmospheric,whalen2020internal} and influence the internal dynamics of stars \citep{rogers2013internal}. 
Recent studies have shown the significant impact of diapycnal mixing, driven by internal gravity waves, on various climate phenomena. However, accurately resolving the short vertical scales involved -particularly in climate simulations- remains challenging. It highlights the importance of theoretical modeling and further study of internal gravity waves \citep{mackinnon2017climate}.

For strongly dispersive waves, as long as amplitudes are small, the weak wave turbulence theory allows writing a closed kinetic equation for the slow evolution of the averaged spectral energy density \citep{hasselmann1966feynman,zakharov1992kolmogorov,nazarenko2011wave,newell2011wave,galtier2022physics}. The first kinetic equation for 3D internal waves, with rotation, was written by 
\citet{olbers1976nonlinear}. Since then, it has been re-derived using various formalisms and assumptions, see e.g. \citep{pelinovsky1977weak,muller1986nonlinear,caillol2000kinetic,lvov2001hamiltonian,lvov2004hamiltonian,lvov2012resonant,scott2024evolution,labarre2024kinetics}. At zero rotation, by applying the hydrostatic approximation -which assumes long horizontal scales compared to vertical scales- several authors found a formal steady spectrum \citep{pelinovsky1977weak,caillol2000kinetic,lvov2001hamiltonian}. But this spectrum is not a physically realizable solution, as noted by \citet{caillol2000kinetic}. Later, \citet{lvov2010oceanic,dematteis2021downscale} showed that there is only one bi-homogeneous steady spectrum that yields a converging collision integral. Yet, this 3D theoretical prediction has never been observed directly. Still in the hydrostatic approximation, \citet{lanchon2023energy} have found a solution to a diffusion approximation of the kinetic equation retaining only induced diffusion triads interactions \citep{mccomas1977resonant}. In 2D, using a more general approach, \citet{shavit2024turbulent} found a steady solution of the full kinetic equation without the hydrostatic approximation. Namely, the steady energy spectrum is
\begin{equation}
	\label{eq:ShavitPrediction}
	e(\bk) \propto k^{-3} |\ok|^{-2}
\end{equation}
where $\bk = (k_x,k_z)$ is the wave vector, $k = \sqrt{k_x^2+k_z^2}$ its modulus, $\ok = N k_x/k$ the internal gravity wave frequency, and $N$ is the buoyancy frequency. Changing coordinates, this is the oceanic Garrett-Munk (GM) spectrum \citep{garrett1979internal} in the limit of zero rotation and short vertical scales compared to the ocean depth
\begin{eqnarray}
k^{-3}\left(\ok/N\right)^{-2}dk_{x}dk_{z}&=k_{z}^{-2}\left(\ok/N\right)^{-2}dk_{z}d\ok \propto e_{GM}.
\end{eqnarray}
Despite deviations from the empirical GM spectrum of measured spectra of oceanic internal gravity waves, it is still considered to be a useful description of oceanic internal gravity waves \citep{polzin2014finescale,dematteis2024interacting}. 

Similar to other anisotropic dispersive waves in fluids, such as Rossby and inertial waves, internal gravity waves interact with slow modes, specifically domain vortical and shear modes \citep{smith2002generation, laval2003forced, brethouwer2007scaling, waite2011stratified, remmel2014nonlinear, howland_mixing_2020, rodda2022experimental}. These slow modes present a significant challenge for the current weak wave turbulence description, which does not account for their evolution or interaction with dispersive waves. Their prominence in the energy spectrum complicates the observation of weak wave turbulence in internal gravity waves, both in direct numerical simulations \citep{remmel2014nonlinear} and experiments \citep{lanchon2023internal}. In particular, a well-observed feature of stratified turbulence is layering (see \citet{caulfield2021layering} and references therein), corresponding to an accumulation of energy in horizontal motions (shear and vortical modes) with the formation of well-mixed layers separated by sharp interfaces. If turbulence and stratification are strong enough, the layers' thickness is ``chosen'' by the flow such that $L_z = U/N$, where $U$ is the typical velocity of the flow \citep{billant_self_2001, brethouwer2007scaling}. This phenomenon looks like a self-organised criticality, where the flow organizes itself such that the Richardson number is close to a critical value \citep{caulfield2021layering}, not too far from the linear stability threshold \citep{miles1961stability,howard1961note}.  

One advances several mechanisms to explain the layering in strongly stratified turbulence. \citet{phillips1972turbulence} proposed a simple model for the evolution of the average vertical density profile, illustrating how layers could develop from small perturbations to an initially linear density profile. He showed that if the magnitude of the buoyancy flux is a decreasing function of the local gradient Richardson number, small disturbances to the initial stratification profile can grow in time. \citet{balmforth1998dynamics} used a reduced model based on two coupled partial differential equations for the average turbulent kinetic energy and the mean buoyancy, and a mixing length model. \citet{taylor2017multi} have reformulated the conditions for amplifying small perturbations to a uniform stratification first proposed by Phillips and introduced a criterion for the development of layering in terms of the spatial distribution of appropriate eddy diffusivity. \citet{petropoulos2023turbulent} used reduced-order models for the evolution of velocity and density gradients and analysed layering in stratified and sheared turbulent flows. They determined the ranges of bulk Richardson numbers and turbulent Prandtl numbers for layering. \citet{billant_experimental_2000,billant_theoretical_2000} showed that vortical modes are prone to zigzag instability, which leads to layering on the scale of $U/N$. Interestingly, one observes the layering in 2D \citep{smith2001numerical}, as well as 3D without vortical modes \citep{remmel2014nonlinear}, or without vortical modes and shear modes \citep{calpelinares_numerical_2020, labarre2024internal}. Also, \citet{fitzgerald2018statistical} showed that stochastic structural stability theory \citep{farrell2003structural} and quasi-linear turbulence closures reproduce layering in the 2D case. Therefore, it is worth searching for a weak wave turbulence explanation of the layering process.

In the present study, we employ 2D DNS of the Boussinesq equations, removing slow (shear) modes. Our prime goal is to identify the parameter regime for observing weak wave turbulence, report fundamental features of the fields in the weak wave turbulence regime - mainly the energy spectrum and the associated energy flux and compare these to the theoretical predictions. We remove shear modes to reach the steady state faster and have a cleaner comparison to weak wave turbulence predictions. The reason to consider 2D flows is twofold – from the practical point of view it is much cheaper and faster compared to 3D DNS. From the theoretical point of view, the kinetic equation for 2D \citep{shavit2023role}, takes a simpler form compared to 3D due to the existence of an additional invariant and has a theoretical prediction of the spectrum outside the hydrostatic limit. Despite being an important simplification when compared to real 3D flows, 2D stratified flows capture important aspects of stratified flows and often serve as an important step towards understanding the 3D problem \citep{smith2001numerical,boffetta_flux_2011,fitzgerald_statistical_2018,calpelinares_numerical_2020}. Also, a 2D description of internal gravity waves is practically relevant both for experiments, e.g. in long water tanks, and in the ocean in the case of internal tides radiated away from isolated 1D topography structures such as the Hawaiian ridge \citep{smith2003tidal}. In the weak wave turbulence regime, excluding low frequencies, we observe a good agreement with the recent theoretical weak wave turbulence prediction \citep{shavit2024turbulent}. As layering kicks in with spectral peaks observed at slow frequencies, our work emphasizes the limitation of the kinetic approach. Yet, our work advances beyond the weak interaction regime, and gives a simple explanation for layering (with scalings for $U$ and $L_z$ based on control parameters) even beyond the weakly non-linear regimes. 

The remaining parts of the manuscript are as follows. In section \ref{sec:Equations}, we introduce the dynamical equations and our notations. In section \ref{sec:Regimes}, we use wave turbulence theory to identify the three regimes: discrete wave turbulence, weak wave turbulence, and strong nonlinear interaction. This emphasizes that to observe the weak wave turbulence regime in a stratified fluid a directed study must be done, which is crucial for experiments and future numerical investigations. We describe our numerical simulations in section \ref{sec:Simulations}, and we analyse them in section \ref{sec:Study}. We present vorticity fields for different stratification and viscosity in subsection \ref{subsec:Vorticity}. Then, we show the 1D energy spectra for the regimes in subsection \ref{subsec:1DSpectra}. In subsection \ref{subsec:SNL}, we analyse in more detail the 2D energy spectra and energy fluxes of a strongly nonlinear simulation. We provide a weak wave turbulence explanation to explain the layering process in subsection \ref{subsec:Layering}. In subsection \ref{subsec:VerificationWWTPrediction}, we analyse the 2D energy spectra and energy fluxes of a weakly nonlinear simulation and compare it to the theoretical predictions of \citet{shavit2024turbulent}. We describe the Doppler shift observed in some of our simulations in subsection \ref{subsec:Doppler}. Section \ref{sec:Conclusion} contains discussions and conclusions.

\section{Governing equations}\label{sec:Equations}

\subsection{Dynamical equations}

Stratified flows can be described most simply using the Boussinesq equations, derived from the Euler equations by assuming a linear density profile in the vertical direction. For two-dimensional flows restricted to the vertical $xz$ plane, the Boussinesq equations are
\begin{align}
	\label{eq:2DBoussinesqIncompressibility}
	\bnabla \cdot \bu &= 0, \\
	\label{eq:2DBoussinesqU}
	\pt \bu + \bu \cdot \bnabla \bu &= - \bnabla p + b \eez  \\  
	\label{eq:2DBoussinesqB}
	\pt b + \bu \cdot \bnabla b &= - N^2 u_z ,
\end{align}
where $x$ and $z$ are respectively the horizontal and vertical coordinates, $\bu = (u_x,u_z)$ is the velocity field, $p$ the kinematic pressure, $b$ the buoyancy, and $N$ is the constant buoyancy (or Brünt-Väisälä) frequency. It is easy to check that (\ref{eq:2DBoussinesqIncompressibility}-\ref{eq:2DBoussinesqB}) have two exact quadratic invariants: the total energy $E=\frac{1}{2}\int\! \diff x  \diff z \left(\bu\cdot\bu+\frac{b^{2}}{N^{2}}\right)$ and the correlation between vorticity, $\nabla^{\perp}\!\times\bu$, and the buoyancy, called  pseudomomentum $P=-\int\! \diff x \diff z (\nabla^{\perp}\!\times\mathbf{u})b$. We consider a periodic domain $\boldsymbol{x} = (x,z) \in \left[0,L\right]^2$ and expand the fields in terms of linear wave modes with wave vector $\bk\in (2\pi\mathbb{Z}/L)^2$. In polar coordinates, $\bk=k(\cos\theta,\sin\theta)$, the dispersion relation is
\begin{equation}\label{eq:omega}
\ok= N \cos\theta. 
\end{equation}
Since the nonlinearity is quadratic, a resonant interaction of internal gravity waves is a triad $(\bk,\boldsymbol{p},\boldsymbol{q})$ satisfying 
\begin{equation}\label{resonant}
\ok\pm\omega_{\boldsymbol{p}}\pm\omega_{\boldsymbol{q}}=0.
\end{equation}   
We set $L=2\pi$, so the $x,z$ components of the wave vectors are integers.

\subsection{Forcing, dissipation and energy transfers}

We add forcing and dissipation to the dynamical equations to study the statistics of stationary
turbulent states. Specifically, we force the system (\ref{eq:2DBoussinesqIncompressibility}-\ref{eq:2DBoussinesqB}) at small wave vectors and add significant dissipation at large wave vectors. Rewriting the dynamical equations in terms of the stream function $\psi$, $(u_x,u_z)=(-\pz\psi,\px\psi)$, with forcing and dissipation yields
\begin{align}
	\label{eq:2DBoussinesqSF}
	\pt \Delta \psi + \left\{ \psi,\Delta\psi \right\} &= \px b + \Delta f_\psi + (-1)^{n-1} \nu_n \Delta^n (\Delta \psi),  \\  
	\label{eq:2DBoussinesqBb}
	\pt b + \left\{ \psi,b \right\} &= -N^2 \px \psi + f_b + (-1)^{n-1}\kappa_n \Delta^n b.
\end{align} 
Here $-\Delta \psi$ is the vorticity, $\left\{ g,f\right\}=\px g \pz f-\pz g \px f$. $f_\psi$ and $f_b$ are respectively the stream function and buoyancy forcing. $\nu_n$ is the hyperviscosity and $\kappa_n$ the hyperdiffusivity. In this study, we set $\kappa_n=\nu_n$. One recovers the standard Boussinesq equations for $n=1$. Using hyperviscosity and hyperdiffusion is a standard method to enlarge the inertial range at a given resolution \citep{brethouwer2007scaling}. 

In Fourier space, equations (\ref{eq:2DBoussinesqSF}-\ref{eq:2DBoussinesqBb}) read 
\begin{align}
	\pt \hpsi - \frac{1}{k^2} \widehat{\left\{ \psi, \Delta \psi \right\}}_{\bk} &= i \frac{k_x}{k^2} \hb + \hfpsi + (-1)^{n-1} \nu_n (-k^2)^n \hpsi \\
	\pt \hb + \widehat{\left\{ \psi, b \right\}}_{\bk} &= i k_x N^2 \hpsi + \hfb + (-1)^{n-1} \kappa_n (-k^2)^{n} \hb
\end{align}
where $(\hat{\cdot})_{\bk}$ denotes the Fourier transform. It follows that the equations for the kinetic energy spectrum $\ekin(\bk)=\left\langle \left| \hbu \right|^2 \right\rangle/2=k^2 \left\langle \left| \hpsi \right|^2 \right\rangle/2$ and the potential energy spectrum $\epot(\bk)=\left\langle \left| \hb \right|^2 \right\rangle/(2N^2)$ are
\begin{align}
	\label{eq:EnergyBudgetKin}
	\pt \ekin(\bk) = \Tkin(\bk) - \Im \mean{k_x \hb \hpsi^*} + \Re \mean{k^2 \hfpsi \hpsi^*} + (-1)^{n-1} \nu_n (-k^2)^n \ekin(\bk), \\
	\label{eq:EnergyBudgetPot}
	\pt \epot(\bk) = \Tpot(\bk) + \Im \mean{k_x \hb \hpsi^*} + \frac{1}{N^2} \Re \mean{\hfb \hb^*} + (-1)^{n-1} \kappa_n (-k^2)^n \epot(\bk),
\end{align}
where $\left\langle \cdot \right\rangle$ is an ensemble average (here a time average in statistically steady state), $(\cdot)^*$ is the complex conjugate, $\Re(\cdot)$ is the real part, $\Im(\cdot)$ is the imaginary part, and the kinetic and potential energy transfers are defined by
\begin{equation}
	\Tkin(\bk) = \Re \left\langle \hpsi^* \widehat{\left\{ \psi, \Delta \psi \right\}}_{\bk} \right\rangle ~~~~ \text{and} ~~~~ \Tpot(\bk) = - \Re \left\langle \hb^* \widehat{\left\{ \psi, b \right\}}_{\bk} \right\rangle 
\end{equation}
respectively. Physically, $\Tkin$ and $\Tpot$ represent the kinetic and potential energy transfers to mode $\bk$ through nonlinear interactions per unit time. We denote the total energy spectrum $e(\bk) = \ekin(\bk) + \epot(\bk)$, the total energy transfer $\mathcal{T}(\bk) = \Tkin(\bk) + \Tpot(\bk)$. We note the 1D kinetic energy transfers as
\begin{align}
	\label{eq:Transferk}
	\Tkin(k) &= \sum\limits_{\bk', |\bk'|\leq k} \Tkin(\bk'),  \\
	\label{eq:Transfertheta}
	\Tkin(\ok) &= \sum\limits_{\bk', |\ok'|\leq \ok} \Tkin(\bk').  
\end{align} 
They represent respectively the kinetic energy transfer to modes with wave vector modulus less than $k$ and the kinetic energy transfer to modes with wave frequency less than $\ok$. We use similar definitions for the 1D potential and total energy transfers. \\

We force the flow with two independent white noise terms for the stream function and the buoyancy for modes such that the forcing amplitudes are non-zero on a bounded ring $|\bk| \in [\kfmin, \kfmax]$ and $k_x,k_y \neq 0$. Without considering other terms but forcing, the dynamical equation for the stream function of forced modes reads
\begin{equation}
	\label{eq:forcing}
	\diff \hpsi = \hat{f}_{\psi,\bk} ~ \diff t = \sqrt{\varepsilon \diff t} ~ \frac{X_{\bk} + X^*_{-\bk}}{\sqrt{ \sum\limits_{\bk} \left| X_{\bk} + X^*_{-\bk} \right|^2 k^2 } }
\end{equation}
where $X_{\bk}$ are complex numbers whose real and imaginary parts are normal random variables. It ensures a real forcing in physical space with a constant average kinetic energy injection rate $\sum\limits_{\bk} \left|\hat{f}_{\psi,\bk} ~ \diff t ~ k \right|^2 /(2\diff t)  = \varepsilon/2$. The equivalent of equation (\ref{eq:forcing}) for the buoyancy is $\mathrm{d} \hb = \hat{f}_{b,\bk} ~ \diff t $ where $\hat{f}_{b,\bk}$ is obtained in the same way than $\hat{f}_{\psi,\bk}$, up to a prefactor $-Nk$, and using an independent stochastic process. Doing so, the average potential energy injection is equal to $\sum\limits_{\bk} \left| \frac{\hat{f}_{b,\bk}}{N} ~ \diff t \right|^2 /(2\diff t)  = \varepsilon/2$. The independence of $f_\psi$ and $f_b$ ensures that the pseudomomentum is zero on average, $\mean{P}=0$. The parameters that quantify the forcing are therefore the average energy injection rate $\varepsilon$, and the wave vector modulus at the middle of the forcing ring $\kf=(\kfmax+\kfmin)/2$.

\section{Parametric regimes}\label{sec:Regimes}

The standard dimensionless parameters for a stratified fluid are the Froude number, which quantifies stratification strength, and the (hyper-viscous) Reynolds number, representing the balance between dissipation and non-linear interaction:
\begin{equation}
	\label{eq:FroudeReynolds}
	Fr \equiv \frac{U}{NL} ~~~~ \text{and} ~~~~ Re_n \equiv \frac{UL^{2n-1}}{\nu_n},
\end{equation}
where $U$ is a typical velocity. We use $U=(\varepsilon/\kf)^{1/3}$, which differs from root mean square velocity used in many other studies \citep{billant_self_2001,brethouwer2007scaling}. The Froude and Reynolds numbers determine fundamental flow length scales: the buoyancy and viscous scales. These are associated with absolute wave vectors:
\begin{equation}
	\label{eq:kbkd}
	\kb \equiv \frac{N}{U} = \frac{Fr^{-1}}{L} ~~~~ \text{and} ~~~~ \kd \equiv \left(\frac{U}{\nu_n}\right)^{1/(2n-1)} = \frac{Re_n^{1/(2n-1)}}{L}.
\end{equation}
For scales larger than $2\pi/\kb$, linear terms dominate over nonlinear ones in the dynamical equations (\ref{eq:2DBoussinesqIncompressibility}-\ref{eq:2DBoussinesqB}), while at scales smaller than $2\pi/\kd$, dissipative terms dominate over nonlinear terms. The third central wave vector corresponds to the box size $2\pi/L$. 
We now identify three regimes: discrete wave turbulence, weak wave turbulence, and strong turbulence. To observe weak wave turbulence,
\begin{enumerate}
  \item Wave amplitudes must be small to ensure weak nonlinear interactions;
  \item The number of pseudo-resonances -meaning resonances nearly satisfying (\ref{resonant})- needs to be large to ensure pseudo-continuous energy exchanges between waves \citep{bourouiba2008discretness,lvov_discrete_2010,buckmaster2021onset}. 
\end{enumerate}
Condition (i) is achieved if dissipation occurs at scales larger than the buoyancy scale, implying
\begin{equation}
	\label{eq:chiw}
	\kd \lesssim \kb ~~ \Rightarrow ~~ Re_n \lesssim Fr^{-2n+1}.
\end{equation}
Therefore, for weak wave turbulence to occur, the Reynolds number cannot be arbitrarily large, as shown in prior studies on internal wave turbulence \citep{lereun2017inertial,lereun2018parametric,brunet2020}. When condition \eqref{eq:chiw} is not met, wave breaking is likely, leading to strongly non-linear stratified turbulence. To ensure condition (ii), the frequency of the nonlinear interaction $\omega_{\rm nl}$ must exceed the wave frequency gap in discrete Fourier space $|\bnabla_{\bk} \ok \cdot \diff\bk|$ with $\diff\bk$, being the wave vector gap between adjacent modes \citep{lvov_discrete_2010}. Using dimensional analysis, we estimate $\omega_{\rm nl} = (\varepsilon k^2)^{1/3}$ and take $\diff\bk=(2\pi s_x /L,2\pi s_z /L)$, leading to:
\begin{equation}
	\label{eq:DiscreteWaveTurbulence}
	\frac{2 \pi N}{L k} |\sin \theta| \sqrt{(s_x \sin \theta)^2 + (s_z \cos \theta)^2} \lesssim \left( U^3 \kf k^2 \right)^{1/3}.
\end{equation}
The prefactor $|\sin \theta|$, implies that the inequality is more likely to be violated at $\theta \simeq \pm \pi/2$, i.e. for small wave frequencies $\ok=N\cos \theta$. Ensuring this condition across all angles $\theta$ and $s_x, s_z =0,\pm 1$ reduces to 
\begin{equation}
	\label{eq:kc}
	k \gtrsim \frac{1}{L} (2\pi)^{3/5} (L \kf)^{-1/5} Fr^{-3/5} \equiv (2\pi)^{3/5} (L \kf)^{-1/5} \kc,
\end{equation}
where
\begin{eqnarray}
	\label{eq:kcbis}
	\kc \equiv Fr^{-3/5}/L. 
\end{eqnarray}
For a constant $L\kf \sim 1$, we expect the interaction to be concentrated on discrete sets of modes with slow wave frequencies and wave vectors modulus $k\lesssim \kc$, leading to discrete wave turbulence \citep{lvov_discrete_2010}. For observing a weak wave turbulence range in the energy spectra, one needs to satisfy
\begin{equation}
	\label{eq:chic}
	\kd \gtrsim \kc ~~ \Rightarrow ~~ Re_n \gtrsim \left[ (2\pi)^{3/5} (L \kf)^{-1/5} \right]^{2n-1} Fr^{-(6n-3)/5},
\end{equation}
Otherwise, the fluid is expected to be in a discrete wave turbulence regime. This observation is crucial for laboratory experiments interested in the weak wave turbulence regime. As our numerical results show in the next section, $\kc$ plays an important role in strongly stratified turbulence. We define $\kc$ through dimensional analysis, so we expect a numerical prefactor of order unity when comparing it to observations. Analogous conditions to equations \eqref{eq:chiw} and \eqref{eq:chic} are also necessary to make comparisons with weak wave turbulence theory for quantum fluids \citep{zhu2022testing}, surface waves \citep{falcon2022experiments}, and other systems. 

In stratified turbulence, the Ozmidov wave vector
$\ko \equiv \sqrt{N^3/\varepsilon}= (L \kf)^{-1/2} Fr^{-3/2}/L$, where the nonlinear time matches $N$, and the Kolmogorov wave vector $\keta \equiv (\varepsilon/\nu_n^3)^{1/(6n-2)} = (L \kf)^{1/(6n-2)} Re_n^{3/(6n-2)}/L$ are often used. For our simulations, we found it convenient to use $\kd$ instead of $\keta$ to set the hyper-viscosity for having well-resolved simulations. Yet, $\kd/\keta = (L \kf)^{-1/(6n-2)} Re_n^{1/(2n-1)(6n-2)}$ so $\keta \simeq \kd$ over the parameter range we investigated. Thus, using $\keta$ instead of $\kd$ as the dissipative scale would not significantly alter conditions \eqref{eq:chiw} and \eqref{eq:chic}. 

Finally, our analysis assumes that $U$ is s determined solely by the input parameters $\varepsilon$ and $\kf$. The observed deviations of the rms velocity in our simulations are discussed in section \ref{sec:Study}. In this section, we also connect our ``naive'' definitions to standard definitions used in studies of stratified flows, using an approximation of the rms velocity based on the discrete wave regime.

\section {Simulations}\label{sec:Simulations}

\subsection{Numerical methods}\label{subsec:NumericalMethods}

We perform forced-dissipated DNS of equations (\ref{eq:2DBoussinesqSF}-\ref{eq:2DBoussinesqBb}) in a square periodic domain of size $L=2\pi$ using a pseudo-spectral method with a standard $2/3$ rule for dealising. For the white noise forcing \eqref{eq:forcing} we set 
\begin{equation}
    [\kfmin, \kfmax] = [0.9;5.1] ~~ \Rightarrow ~~ \kf= \frac{\kfmin + \kfmax}{2}=3 ~~~~ \text{and} ~~~~ \varepsilon = 10^{-3}. \\
\end{equation}
For time advancement, we employ the fractional-step splitting method \citep{mclachlan_quispel_splitting_2002}. Namely, we apply the linear operator for a time increment $\diff t/2$, compute the contribution of the nonlinear term using the Runge-Kutta 2 method, and apply the linear operator for $\diff t/2$. This method has the advantage of treating the linear terms explicitly and achieving second-order precision. We denote by $M$ the number of grid points in each direction. The hyperviscosity order is set to $n=4$ and the hyperviscosity $\nu_n$ is fixed such that the ratio between the maximal wave vector modulus $\kmax=M/3$ and the viscous wave vector $\kd$ (\ref{eq:kbkd}) is $1.5$. The time step $\mathrm{d}t$ is the minimum between $10^{-2}/N$ and the time step given by a CFL number $0.65$. These choices allow us to obtain well-resolved simulations for the investigated range of parameters. 

\subsection{Setting up the data set}\label{subsec:Dataset}

To construct our dataset, we first run simulations at a small resolution $M=128$. We start simulations with $M\geq256$ from the end of the simulation at the same $N$ with lower resolution $M/2$ and decrease the viscosity as we increase the resolution. It allows us to save computational time and reach statistically steady states faster. The simulation time of the small resolution simulations, i.e. $M=128$, is $\propto 400 N$. For the other simulations, i.e. $M\geq256$, the simulation time is $\propto 200 N$. Then, our runs are much longer than the kinetic time, which is necessary to reach a statistically steady state of weak internal gravity wave turbulence. We give the list of the simulations, with the values of the relevant parameters, in Tab.~\ref{tab1}. 

\begin{table}
	\centering
	\begin{tabular}{|p{3cm}|p{2.5cm}|p{2.5cm}|}
		\hline
		{$\boldsymbol{M}$} & $\boldsymbol{N}$  \\
		\hline
		$128$ & $1,2,4,8,16,32$ \\
		$256$ & $1,2,4,8,16,32$ \\
		$512$ & $1,2,4,8,16$  \\
		$1024$ & $1,2,4,8,16$ \\
		$2048$ & $1,2,4,8$  \\
		\hline
	\end{tabular}
	\caption{List of our simulations with relevant control parameters. We set $L=2\pi$, $n=4$, $\varepsilon=10^{-3}$, $\kf=3$, and $\kmax/\kd=1.5$. Therefore, the dimensionless parameters (\ref{eq:FroudeReynolds}) are then given by $Fr = 1/[(3000)^{1/3} 2\pi N]$ and $Re_n= (4\pi M / 9)^{7}$.}
	\label{tab1}
\end{table}

\section{Study of the flow regimes}\label{sec:Study}

\begin{figure}
	\centering
	\includegraphics[scale=1]{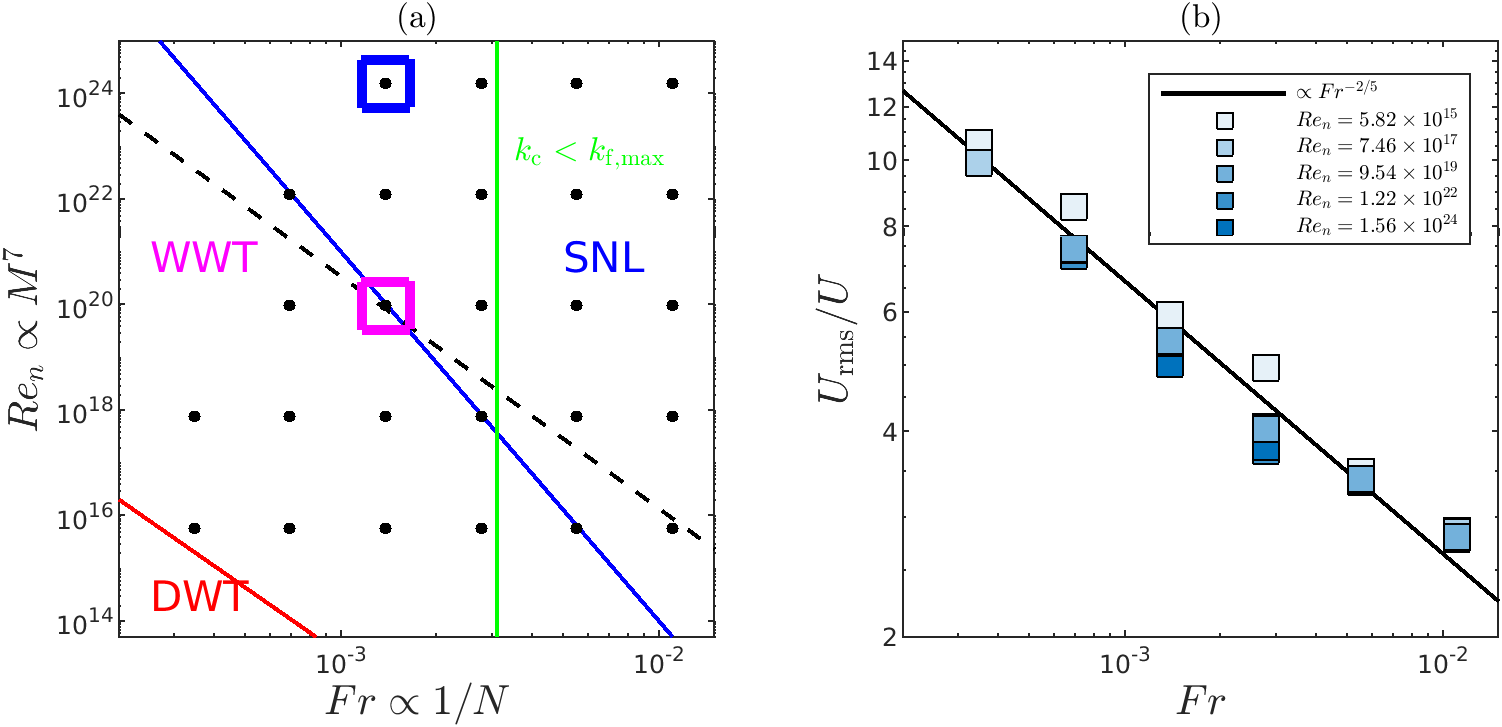} 
	\captionsetup{width=\textwidth}
	\caption{(a) Simulations grid on the parametric plane ($Fr$,$Re_n$). The blue line represents $\kb = \kd$ \eqref{eq:chiw}, which separates the weak wave turbulence from the strong nonlinear regimes. The red line indicates $\kc = \kd$ \eqref{eq:chic}, distinguishing the weak wave turbulence from the discrete wave turbulence regimes. The dashed line corresponds to the transition (\ref{eq:Sweeping}), with $\alpha=10$. The green line represents $\kc = \kfmax$. The blue box highlights the simulation whose spectra are shown in Fig.\ref{fig:spectralSNL}. The magenta box indicates the simulation whose spectra are shown in Fig.\ref{fig:spectralWWT}. (b) Ratio between the r.m.s velocity $U_{\rm rms}$ and the ``naive'' velocity scale $U=(\varepsilon/\kf)^{1/3}$ as a function of $Fr$ for all simulations with varying $Re_n$. The dashed line corresponds to the theoretical scaling (\ref{eq:UL}). \label{fig:phase_space_urms}}
\end{figure}

We present our simulations on the parametric ($Fr$,$Re_n$) plane and the transition lines for the expected regimes in Fig.~\ref{fig:phase_space_urms}(a). In Fig.\ref{fig:phase_space_urms}(b) We plot the ratio between the r.m.s velocity $U_{\rm rms} = \sqrt{2 E_{\rm kin}}$ and the ``naive'' velocity scale $U=(\varepsilon/\kf)^{1/3}$, where $E_{\rm kin}$ is the total kinetic energy. This ratio decreases like $Fr^{-2/5}$, and is weakly dependent on $Re_n$.

\subsection{Vorticity structure across Regimes} \label{subsec:Vorticity}

\begin{figure}
	\centerline{\includegraphics[width=\linewidth]{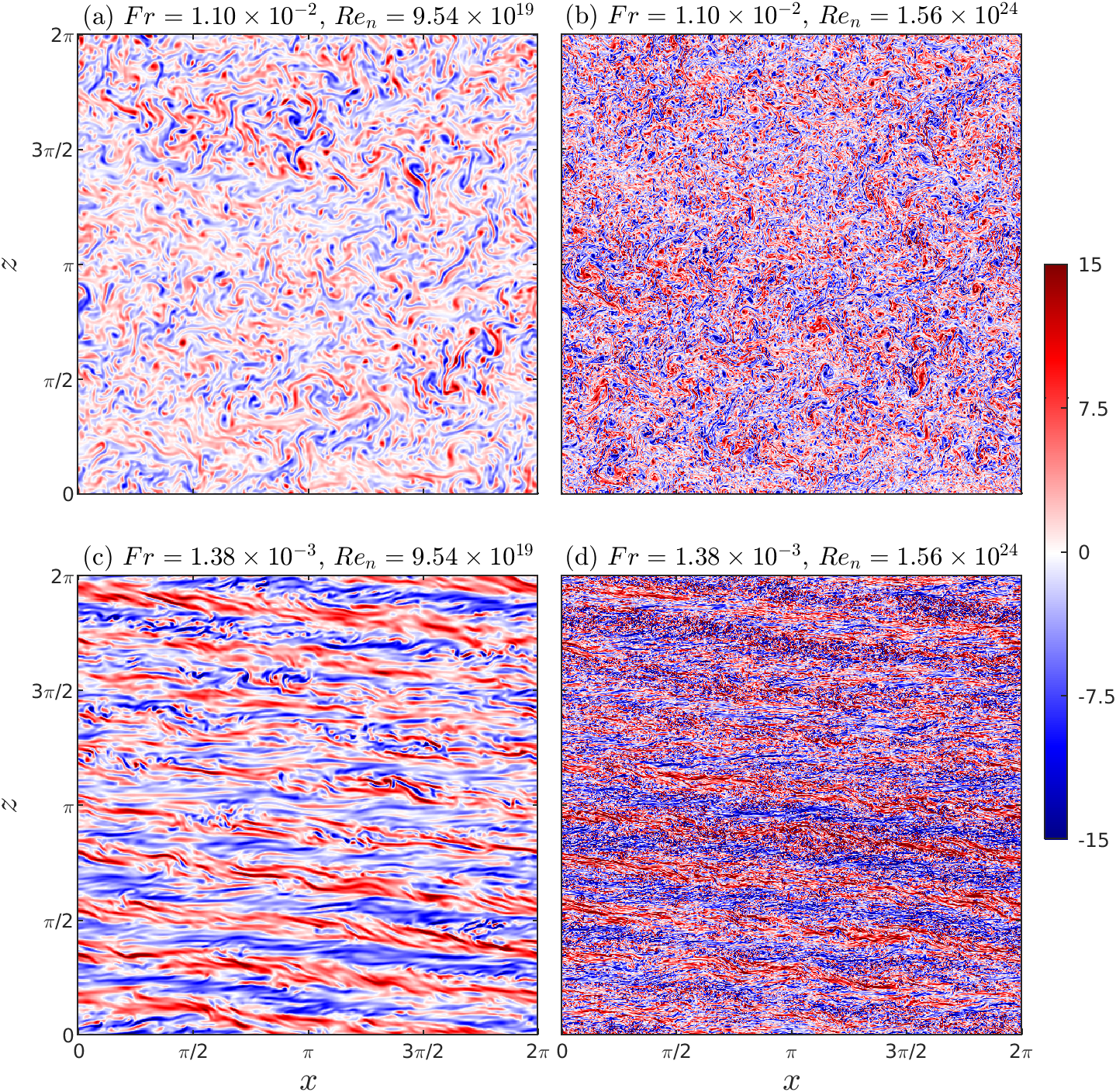}}
    \captionsetup{width=\textwidth}
	\caption{Vorticity field in the statistically steady state for simulations with different $Fr$ and $Re_n$.}
	\label{fig:vorticity}
\end{figure}

In Fig.\ref{fig:vorticity}, we present the vorticity field in the statistically steady state for four simulations. When stratification is weak ($Fr$ is large) and moderate Reynolds number $Re_n$, we observe 2D vortices without layering (Fig.\ref{fig:vorticity}(a)). As $Re_n$ increases, smaller vortices appear (Fig.\ref{fig:vorticity}(b)), as expected due to the extension of the inertial range. With stronger stratification (smaller $Fr$) we observe layering in the vorticity field (Fig.\ref{fig:vorticity}(c)). This phenomenon has been reported previously in simulations without shear modes \citep{calpelinares_numerical_2020}. Notably, the layers' thickness remains unchanged after further increases in $Re_n$; while the vortices become smaller (Fig.\ref{fig:vorticity}(d)).

\subsection{Energy spectra across regimes} \label{subsec:1DSpectra}

\begin{figure}
	\centerline{\includegraphics[width=\linewidth]{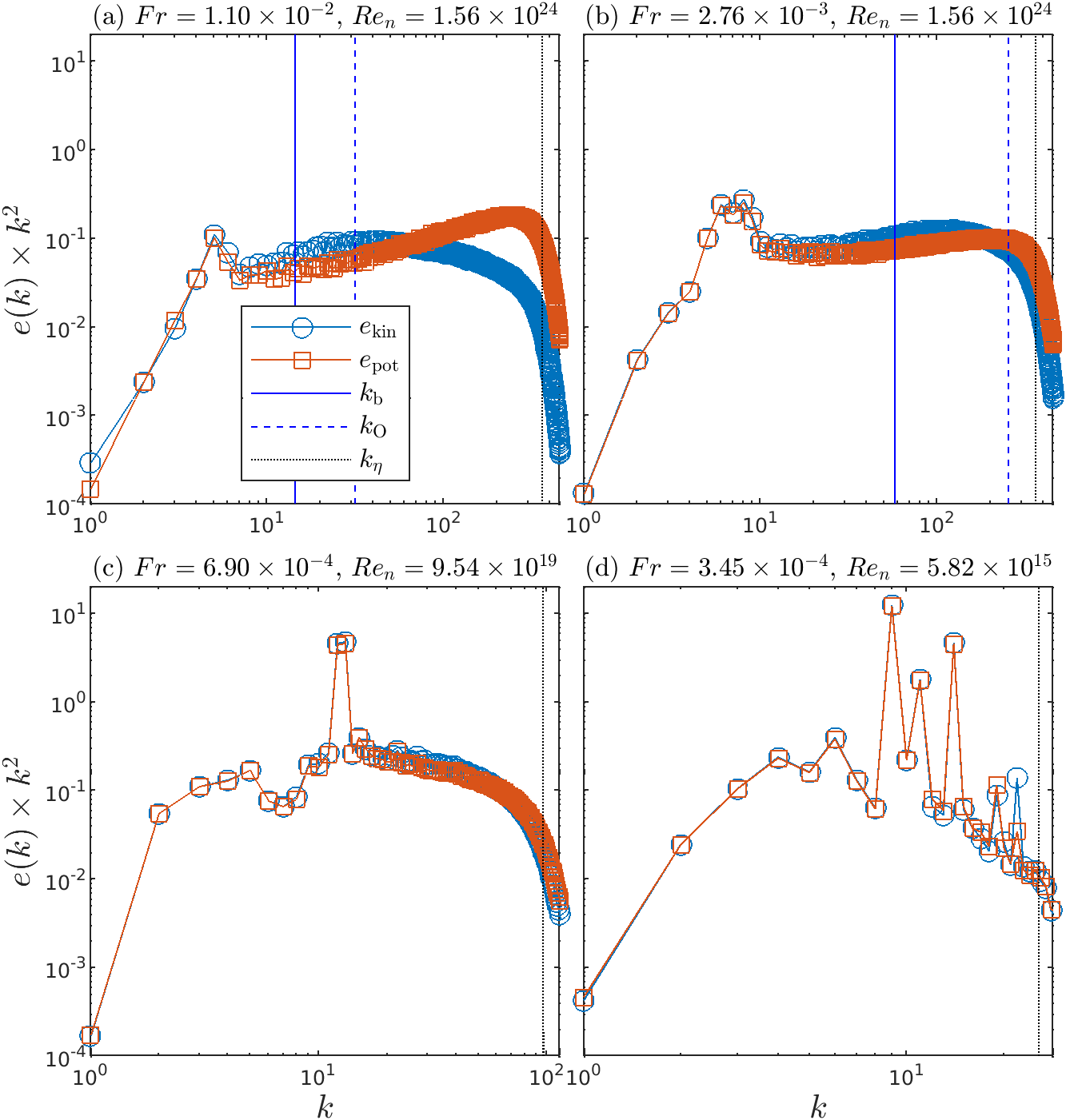}}
    \captionsetup{width=\textwidth}
	\caption{1D kinetic and potential energy spectra for four simulations, compensated by $k^2$. Vertical lines correspond to buoyancy, Ozmidov, and Kolmogorov wave vectors $\kb$, $\ko$, and $\keta$. For each panel, we show only the range $k \in [1:\kd]$, which contains almost all the energy.}
	\label{fig:1Dspectra}	
\end{figure}

In Fig.\ref{fig:1Dspectra}, we present the compensated 1D spectra
\begin{eqnarray}
    e(k) = \sum\limits_{\bk', k-1\leq k'<k} e(\mathbf{k})
\end{eqnarray}
for four simulations across various regimes: (a) strong nonlinear - weakly stratified regime, (b) strong nonlinear - strongly stratified regime, (c) weak wave turbulence, and (d) discrete wave interaction. 

When the stratification is weak and the nonlinearity is strong (a), the potential energy spectrum differs from the kinetic energy spectrum, particularly for $k>\ko$. In this case, the spectra are continuous, except at the end of the forcing range. As stratification increases (b), both $\kb$ and $\ko$ become larger,  leading to closer alignment between the kinetic and potential energy spectra across a broader range. In this simulation, nearly all energetic scales are influenced by stratification, as $\ko \simeq \keta$. In the weak wave turbulence regime (c), characterised by intermediate stratification and nonlinearity, we satisfy condition (\ref{eq:chiw}) so all energetic scales are expected to interact through weak nonlinearity. The potential and kinetic energy spectra are nearly equal across all scales, which is typical for internal gravity waves. Yet, the energy spectrum peaks around $k=13$, with only the range $k \gtrsim 13$ appearing continuous. The peak, which corresponds to the layering, also perturbs the scaling for large wave vectors $k \geq 13$ and we do not observe the theoretical scaling in this simulation. For the simulation in panel (d), the stratification is the highest, and nonlinearity is the weakest such that condition (\ref{eq:chic}) is almost violated. In this regime, energetic scales interact predominantly through a discrete set of interactions, as evidenced by the numerous discrete peaks in the spectrum. It confirms that the simulation corresponds to the discrete wave interaction regime.

\subsection{Strong non linearity regime} \label{subsec:SNL}

\begin{figure}
	\centerline{\includegraphics[width=\linewidth]{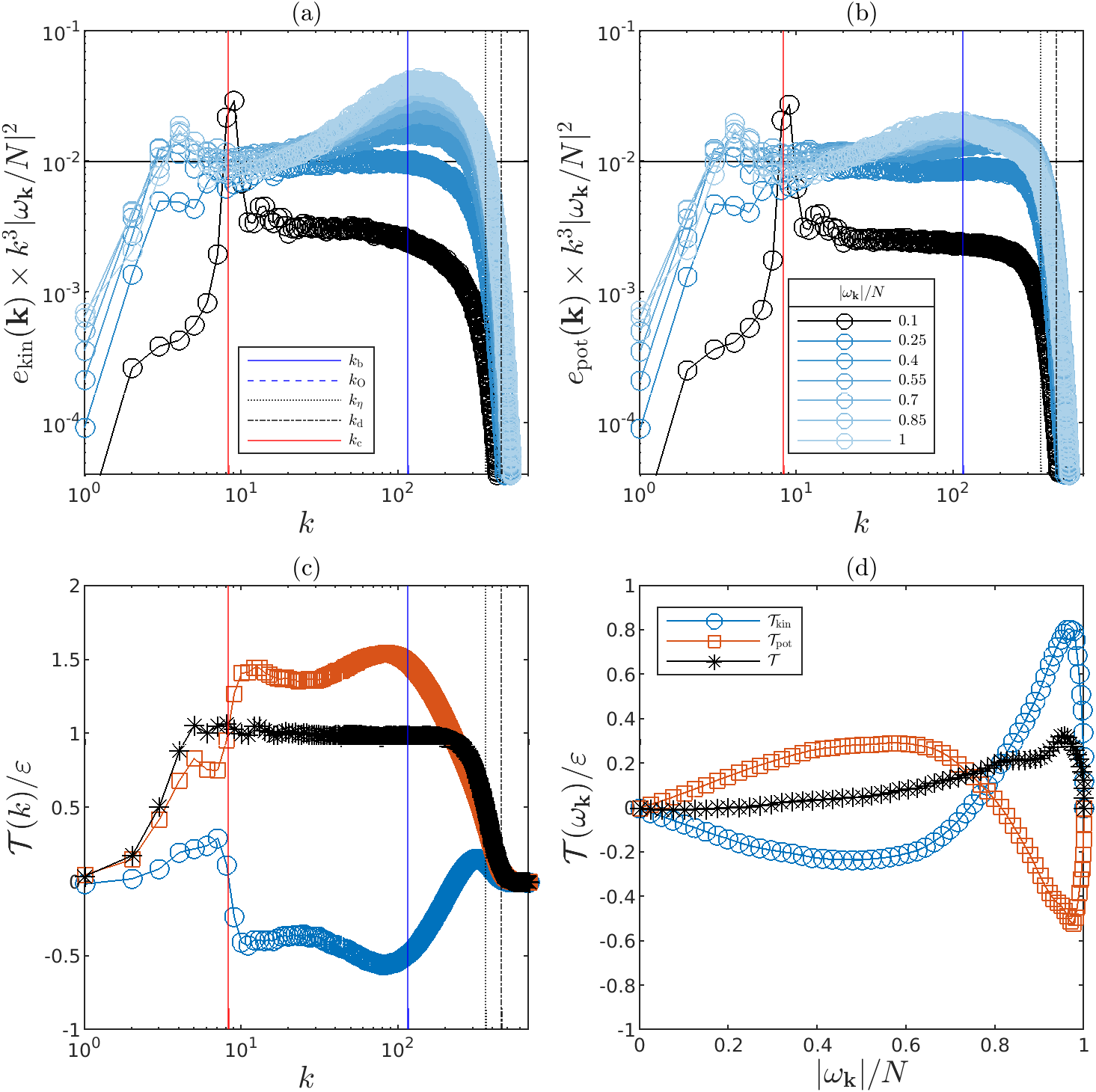}}
    \captionsetup{width=\textwidth}
	\caption{Energy spectra and energy transfers for the simulation 
    in the strong nonlinear regime with $Fr=1.38 \times 10^{-3}$ and $Re_n=1.56 \times 10^{24}$. (a) Slices of the compensated kinetic energy spectrum and (b) slices of the compensated potential energy spectrum for different wave frequencies. (c) Normalised energy transfers as a function of $k$. (d) Normalised energy transfers as a function of $|\ok|/N$. Legend in panel (a) is used for panels (b) and (c), legend in panel (b) is used for panel (a), and legend in panel (d) is used for panel (c).}
	\label{fig:spectralSNL}
\end{figure}

In Fig.\ref{fig:spectralSNL}(a-b), we show slices of the energy spectra at various frequencies as a function of the wave vector amplitude $k$, compensated by the weak wave turbulence prediction (\ref{eq:ShavitPrediction}), for a simulation in the strong nonlinearity regime. We see in panel (a) that the kinetic energy spectrum is shallower than $k^{-3}$ and has a maximum at $k=\kb$ and high $\ok$, as observed in earlier studies (see e.g. \citet{waite2011stratified,augier_stratified_2015}). For slow frequencies $\ok/N = 0.1$ we observe a peak at the critical wave vector $k \simeq \kc$ corresponding to layering. It indicates that $\kc=Fr^{-3/5}/L$, obtained using weak wave turbulence theory in section \ref{sec:Regimes}, is relevant for strongly stratified flows beyond the weakly nonlinear regime. The potential energy spectrum, shown in panel (b), follows the same trends with a less pronounced maximum at $k=\kb$. 

In panel (c), we show the energy transfers (\ref{eq:Transferk}):
\begin{enumerate}
	\item For $k\lesssim \kc$, the transfers $\Tkin$ and $\Tpot$ have no clear behavior, with a sharp transition around $\kc$;
	\item For $k \in [\kc,\kb]$, the transfers have clear tendencies. Potential energy goes forward (to small scales) while converted to kinetic energy, and the kinetic goes backward (to large scale) and is converted back to potential energy. The total energy transfer equals the average injection rate $\varepsilon$. Interestingly, this picture is consistent with the energy cycle explained in \citet{muller1986nonlinear} (see Figs.27 and 28 of this reference).
	\item In the range $k \in [\kb,\keta]$, $\Tpot$ decreases and $\Tkin$ increases such that both are positive, and their sum is still $\varepsilon$.
	\item At $k\simeq \keta$, $\Tkin$ and $\Tpot$ decrease to zero. The transfers are negligible only when $k \simeq \kd$, and it is tempting to interpret the range $k\in [\keta,\kd]$ as an ``intermittent'' range. Yet, the smallness of this range does not allow it to be conclusive.
\end{enumerate}
In panel (d), we show the energy transfers (\ref{eq:Transfertheta}) as a function of $|\ok|/N = |\cos \theta|$. $\Tkin$ and $\Tpot$ are non-monotonous and have opposing trends. At low wave frequencies, the kinetic energy goes to smaller $\ok$ and the potential energy to larger $\ok$. It explains the layering observed in the vorticity field for slow waves at low $Fr$ (Fig.\ref{fig:vorticity}(c-d)). The total energy transfer is positive for all wave frequencies meaning that the total energy transfer is toward higher $\ok$.

\subsection{Layering} \label{subsec:Layering}

We explain the layering in our strongly stratified simulations as follows: The potential energy goes forward in scale and is converted into kinetic energy that goes backward in scale; The backward kinetic energy cascade stops at $k\simeq \kc$ due to the discreteness of the wave-wave interactions so energy accumulates at this scale; Since the condition (\ref{eq:DiscreteWaveTurbulence}) is more easily broken at small wave frequencies, we expect the energy accumulation to be for small $\ok$. 

If the inverse energy transfers stop at $k \simeq \kc$, a mechanism must act against the accumulation of kinetic energy. Otherwise, simulations of 2D stratified turbulence without shear modes would not reach a statistically steady state, as observed here and in \citet{calpelinares_numerical_2020}. One possible mechanism is that the energy carried by the inverse kinetic transfers is converted to potential energy at $k \simeq \kc$. An order of magnitude estimate of equations (\ref{eq:2DBoussinesqU}-\ref{eq:2DBoussinesqB}) yields the large scale flow velocity $U_{\rm L}$:
\begin{equation}
	\label{eq:UL}
	|\bu \cdot \bnabla \bu| \sim |b| ~~~~ \Rightarrow ~~~~ U_{\rm L}^2 \kc \sim N  U_{\rm L} ~~~~ \Rightarrow ~~~~ U_{\rm L} \propto \frac{N}{\kc} = U Fr^{-2/5},
\end{equation}
where we have used $|b| \sim N U_{\rm L}$ and (\ref{eq:kcbis}). This mechanism is impossible for shear modes, which have zero vertical velocity. It may explain why strongly stratified simulations with shear modes take a much longer time to reach a steady state \citep{smith2001numerical,smith2002generation} (see also \citep{brethouwer2007scaling} and references therein). When shear modes are present, the steady state is reached after several instabilities \citep{caulfield2021layering}, leading to the increase of the layers' thickness and large-scale flow velocity \citep{remmel2014nonlinear, fitzgerald2018statistical}. We can also obtain the scaling (\ref{eq:UL}) by assuming that the flow reaches a critical Richardson number. Namely, using the estimate $|\pz \mean{u_x}| \sim \kc U_{\rm L}$, we can write
\begin{equation}
	\label{eq:Richardson}
	Ri \equiv \frac{N^2}{ \left( \pz \mean{u_x} \right)^2 } = Ri_c = \text{constant} ~~~~ \Rightarrow ~~~~ U_{\rm L} \propto \frac{N}{\kc} = U Fr^{-2/5}.
\end{equation} 

The two reasoning (\ref{eq:UL}) and (\ref{eq:Richardson}) mean that the vertical Froude number based on $L_z = 2\pi/\kc$ and $U_{\rm L}$, namely $Fr_z^* = U_{\rm L} /(N L_z)$, is of order unity. It is therefore consistent with earlier studies \citep{billant_self_2001, brethouwer2007scaling}. For our simulations, where layering is present, $U_{\rm rms}=\sqrt{2 E_{\rm kin}}$ is close to the large scale flow velocity $U_{\rm L}$. We observe in Fig.\ref{fig:phase_space_urms}(b) that the expected scaling (\ref{eq:UL}) is verified. Therefore, the relation (\ref{eq:UL}) gives a first-order estimate of the typical velocity of 2D stratified flows when shear modes are removed, at least for $Fr$ and $Re_n$ investigated here. 

It allows us to define the turbulent Froude and (hyper-viscous) Reynolds numbers \citep{brethouwer2007scaling}
\begin{equation}
	\label{eq:TurbulentFroudeRenolds}
	Fr^* \equiv \frac{\varepsilon}{N U_{\rm L}^2} = \left( L\kf \right) ~ Fr^{9/5} ~~~~ \text{and} ~~~~ Re_n^* \equiv \frac{U_{\rm L}^{6n-2}}{\nu_n \varepsilon^{2n-1}} = \left( L\kf \right)^{1-2n} ~ Re_n ~ Fr^{(-12n+4)/5}
\end{equation}
that are based on $U_{\rm L} \propto U_{\rm rms}$. Note that the turbulent Froude number is much lower than $Fr$ and the turbulent Reynolds number much larger than $Re_n$ for small $Fr$ and large $Re_n$. We can also compute the associated buoyancy wave vector and the viscous wave vector as follows
\begin{eqnarray}
	\label{eq:Turbulenkbkd}
	\kb^* \equiv \frac{Fr^*}{L} ~~~~ \text{and} ~~~~ \kd^* \equiv \frac{Re_n^{*1/(2n-1)}}{L}.
\end{eqnarray}
Equations (\ref{eq:TurbulentFroudeRenolds}-\ref{eq:Turbulenkbkd}) link our naive definitions (\ref{eq:FroudeReynolds}-\ref{eq:kbkd}) with the definitions widely used in other studies of stratified flows, as far as the scaling (\ref{eq:UL}) is valid.

\subsection{Verification of the weak wave turbulence prediction} \label{subsec:VerificationWWTPrediction}

Weak wave turbulence means that, though non-linearly interacting, the linear waves remain the main degrees of freedom and carry most of the energy \citep{lereun2017inertial,lereun2018parametric,yokoyama2019energy,lam2020partitioning}). It can be easily verified through the spatiotemporal energy spectrum, that is the Fourier transform in time of the energy spectral density:
\begin{equation}
	\label{eq:SpatiotemporalSpectrum}
	e(\bk,\omega) = \frac{k^2 \left| \hpsi(\omega) \right|^2}{2} + \frac{\left| \hb(\omega) \right|^2}{2N^2},
\end{equation}
where $\hpsi(\omega)$ and $\hb(\omega)$ are the Fourier transform in time of $\hpsi(t)$ and $\hb(t)$. Physically, $e(\bk,\omega)$ is the energy density at wave vector $\bk$ and temporal frequency $\omega$. If the flow is of weakly nonlinear waves, we expect $e(\bk,\omega)\sim \delta(\omega \pm \ok)$. Otherwise, strong nonlinear interactions or other flow modes are important \citep{nazarenko2011wave}. 

\begin{figure}
	\centerline{\includegraphics[width=\linewidth]{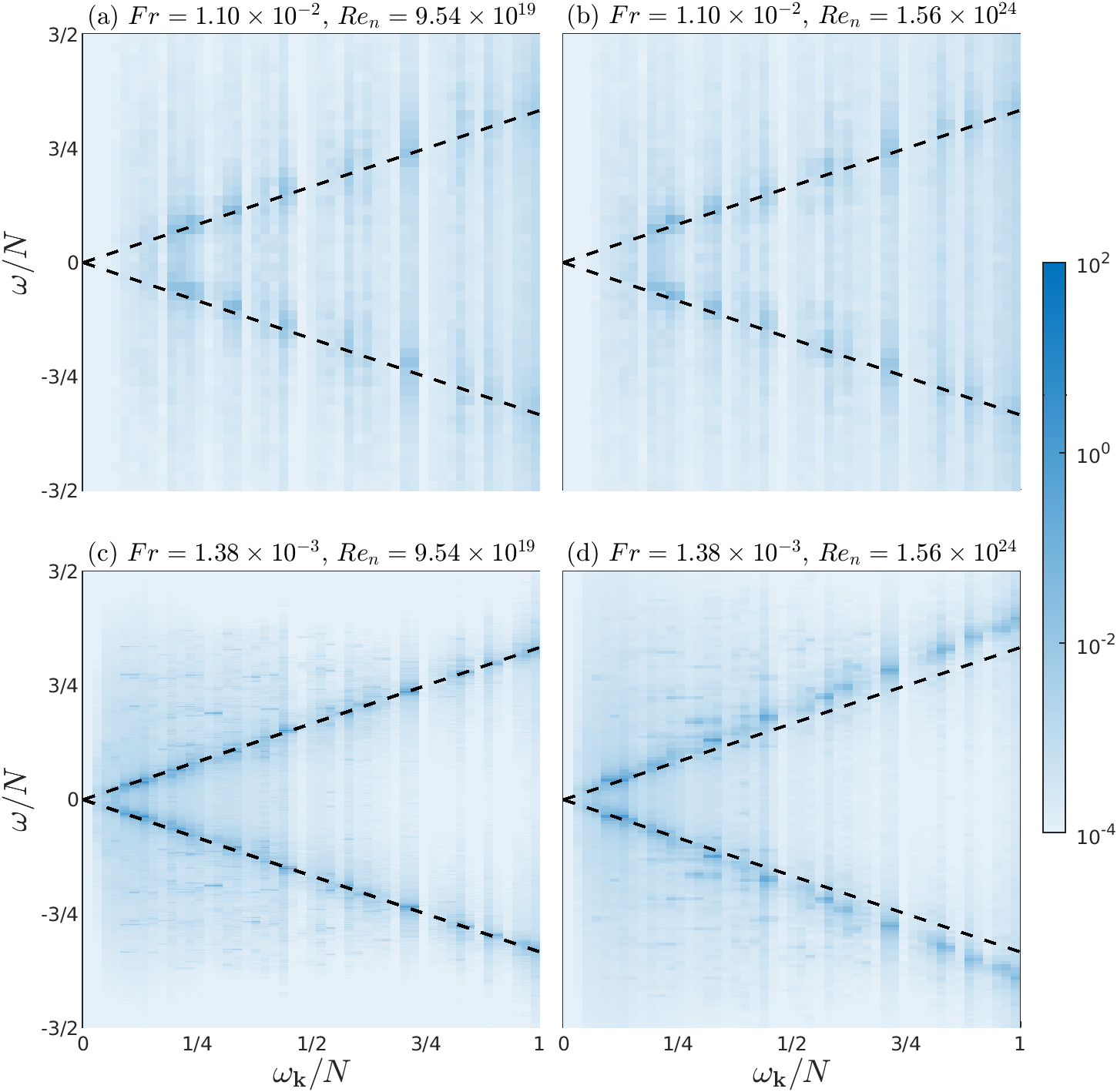}}
	\captionsetup{width=\textwidth}
	\caption{Spatiotemporal energy spectrum $e(\ok,\omega)$ for four of our simulations. In all panels, the dashed lines correspond to the dispersion relation $\omega = \pm \ok$.}
	\label{fig:spatiotemporal}
\end{figure}

In Fig.\ref{fig:spatiotemporal}, we show the spatiotemporal energy spectrum as a function of $\ok$ and $\omega$ for four simulations, obtained after a straightforward summation over $\bk$. To compute this spectrum, we use modes with $|\bk| \leq M/4 < \kmax$ to save computational time. Since these modes contain most of the energy, this truncation has little impact on the results. For weak stratification, the energy is not only on the linear dispersion relation curve, as shown in Fig.\ref{fig:spatiotemporal}(a-b). For a simulation at higher stratification, shown in panel (c), we see that most of the energy is concentrated near the linear dispersion relation, meaning that this simulation is more likely to meet weak wave turbulence assumptions. We will use this simulation to compare the energy spectra to the theoretical prediction. 

\begin{figure}
	\centerline{\includegraphics[width=\linewidth]{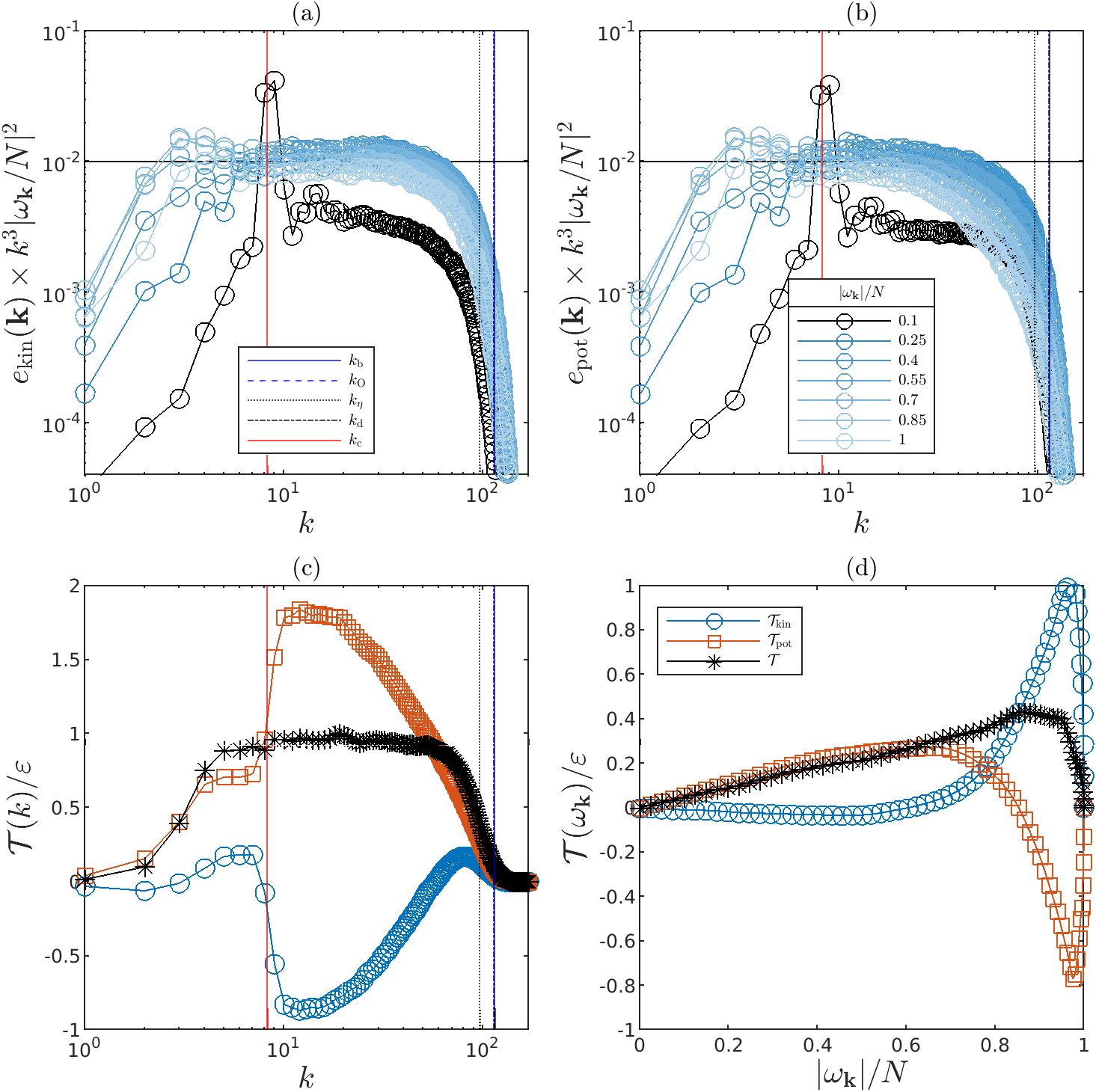}}
    \captionsetup{width=\textwidth}
	\caption{Energy spectra and energy transfers for the simulation with $Fr=1.38 \times 10^{-3}$ and $Re_n=9.54 \times 10^{19}$, in the weak wave turbulence regime. (a) Slices of the compensated kinetic energy spectrum and (b) slices of the compensated potential energy spectrum for different wave frequencies. (c) Normalised energy transfers as a function of $k$. (d) Normalised energy transfers as a function of $|\ok|/N$. Legend in panel (a) is used for panels (b) and (c), legend in panel (b) is used for panel (a), and legend in panel (d) is used for panel (c). \label{fig:spectralWWT}} 
\end{figure}

In Fig.\ref{fig:spectralWWT}(a-b), we show slices of the kinetic and potential energy spectra, compensated by the weak wave turbulence prediction (\ref{eq:ShavitPrediction}), for the simulation with $Fr=1.38 \times 10^{-3}$ and $Re_n=9.54 \times 10^{19}$. We see a good agreement between the theory and our DNS, except for low frequencies $\ok/N$ where the spectral peak due to layering is present. Also, the potential energy spectrum decreases faster than the kinetic energy spectrum, particularly at high $\ok/N$, because the potential energy is ``taxed'' by the conversion to potential energy while transferred to small scales. We see in Fig.\ref{fig:spectralWWT}(c-d) that the energy transfers are similar to the simulation with larger $Re_n$ (Fig.\ref{fig:spectralSNL})(c-d), except that there is less inertial range and $\kd \simeq \kb$ so we do not observed the bump at $k=\kb$. To our knowledge, it is the first verification of the theoretical prediction for weak non-hydrostatic internal gravity waves.

\subsection{Doppler shift} \label{subsec:Doppler}

Keeping stratification high and increasing the nonlinearity compared to the weak wave turbulence simulation, we see that the spatiotemporal spectrum gets a shift from the dispersion relation $\omega = \pm \ok$, visible in Fig.\ref{fig:spatiotemporal}(d). This Doppler shift is known to occur in rotating and/or stratified flows due to the nonlinear interactions with the large-scale flow (see e.g. \citep{dileoni2015absorption,campagne2015disentangling,lam2020partitioning}). To quantify this Doppler shift, we compare the linear wave frequency to the empirical frequency 
\begin{eqnarray}
	\label{eq:EmpiricalFrequency}
	\oemp(\ok) = \left. \sum\limits_{\omega} \omega ~ e(\ok,\omega)  \right/  \sum\limits_{\omega}  e(\ok,\omega).
\end{eqnarray}

\begin{figure}
	\centerline{\includegraphics[width=\linewidth]{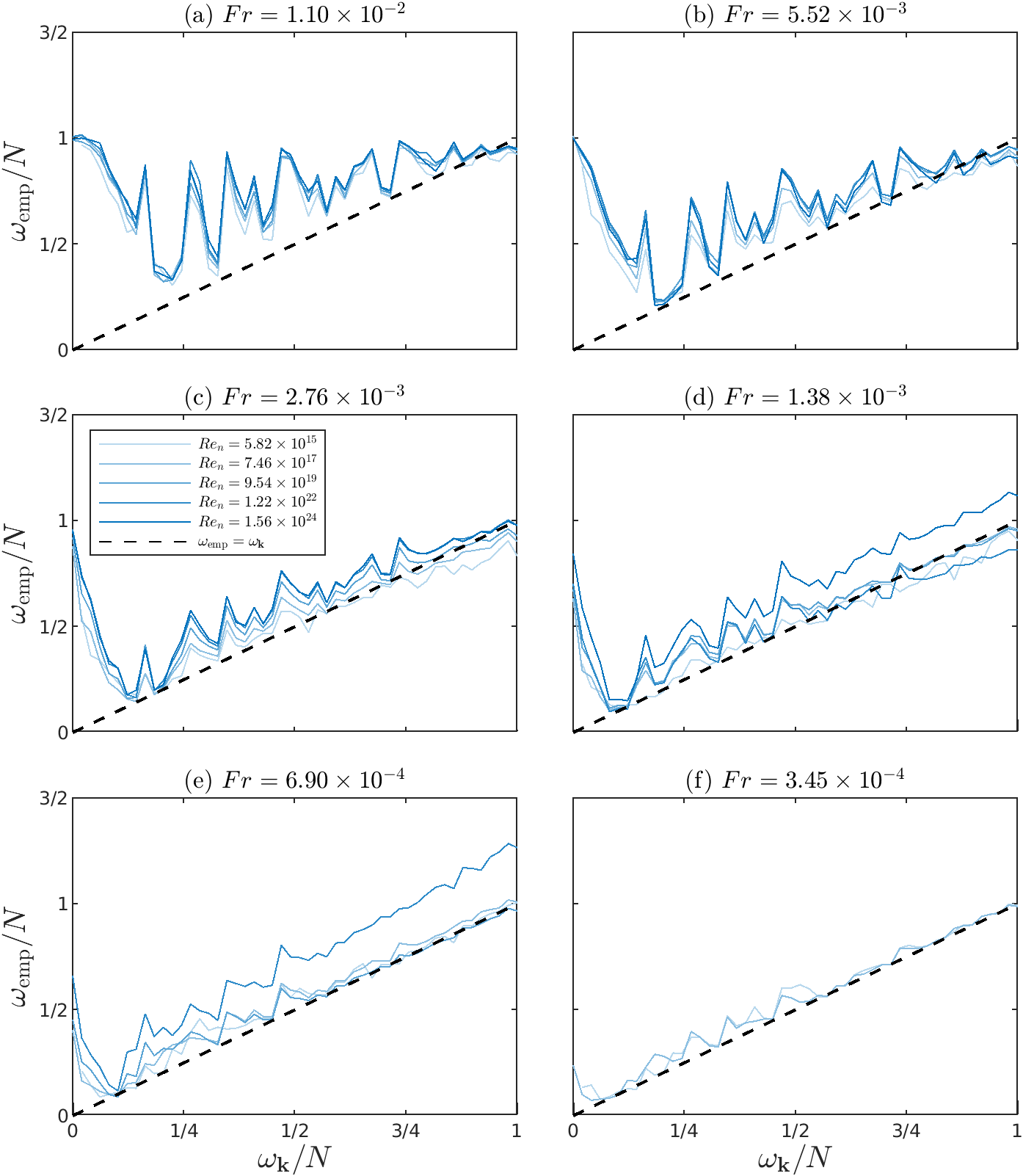}}
	\caption{Empirical frequency (\ref{eq:EmpiricalFrequency}) vs the wave frequency $\ok$ for all our simulations. In all panels, the dashed line indicates $\oemp =\ok$.}
	\label{fig:sweeping}
\end{figure}

In Fig.\ref{fig:sweeping}, we show $\oemp$ as a function of $\ok$ for all our simulations. Panel (a) corresponds to weak stratification $Fr=1.10 \times 10^{-2}$. In this case, $\oemp/N$ is close to unity for all wave frequencies, meaning that energy is not on the linear dispersion relation. The empirical frequency depends weakly on the Reynolds number. In panel (b), for $Fr=5.52 \times 10^{-3}$, $\oemp$ gets closer to the wave dispersion relation for higher $\ok$ and still has a weak dependence on $Re_n$. In panel (c), for $Fr=2.76 \times 10^{-3}$, $\oemp$ is close to $\ok$ except at small wave frequencies. We do not observe the Doppler shift for weak stratification at $Fr > 2.76 \times 10^{-3}$ when $\kc$ is less or close to $\kfmax$ (see Fig.\ref{fig:phase_space_urms}(a)). This observation is in line with \citet{dileoni2015absorption}, who explained that the layering process requires external forcing not to disrupt the development of the large-scale flow. In panel (d-f), for $Fr\leq1.38 \times 10^{-3}$, we observe different behavior depending on the Reynolds number. For lowest $Re_n$, $\oemp$ gets closer to $\ok$ as the increases. On the contrary, for the highest $Re_n$, we observe a significative shift in $\oemp$. The Doppler shift is visible for three simulations: $(Fr=1.38 \times 10^{-3},Re_n=1.22 \times 10^{22})$, $(Fr=1.38 \times 10^{-3},Re_n=1.56 \times 10^{24})$ and $(Fr=6.90 \times 10^{-4},Re_n=1.22 \times 10^{22})$, shown in panels (d-e). It prevented us from using these simulations to compare weak wave turbulence predictions for the energy spectrum. 

It is known that the Doppler shift appears when the sweeping frequency due to the large scale flow becomes larger than the linear frequency (see e.g. \cite{dileoni2015absorption}). For our simulations, it yields to the condition
\begin{equation}
	U_{\rm L} k \gtrsim N.
\end{equation}
If we want it to be valid in the inertial range, namely $k \lesssim \keta \propto Re_n^{3/{6n-2}}/L$, it follows that
\begin{equation}
	\label{eq:Sweeping}
	U Fr^{-2/5} \frac{Re_n^{3/{6n-2}}}{L} \gtrsim \alpha N ~~~~ \Rightarrow ~~~~ Re_n \geq \alpha^{(6n-2)/3} ~ Fr^{(2-6n)/5},
\end{equation}
where we have used (\ref{eq:UL}) and $\alpha$ is an undetermined numerical constant. The physical meaning of (\ref{eq:Sweeping}) is that the mean flow starts to create frequencies bigger than $N$, that are not damped by viscosity. In Fig.\ref{fig:phase_space_urms}(b), we show the line given by condition (\ref{eq:Sweeping}) with $\alpha=10$. The simulations affected by the Doppler shift are above this line and for $\kc > \kfmax$.

\section{Conclusions and discussions}\label{sec:Conclusion}

We performed direct numerical simulations of 2D stratified turbulence without shear modes (also called Vertically Sheared Horizontal Flows). In the weak wave turbulence regime, we verified the theoretical predictions outside the hydrostatic limit \citep{shavit2024turbulent}. The energy spectrum agrees with the theory except for low wave frequencies. Yet, our simulations are subject to layering -an accumulation of energy in slow waves- which perturbs the theoretical predictions. This layering occurs also outside the weakly nonlinear regimes, as observed in many earlier studies \citep{smith2002generation,laval2003forced,waite2011stratified,remmel2014nonlinear,fitzgerald2018statistical,calpelinares_numerical_2020}. 

We explain the layering by the inverse kinetic energy transfers and the discreteness of the wave-wave interactions at large scales, typical of weakly nonlinear wave systems \citep{nazarenko2011wave}. It allows us to obtain quantitative predictions for the layers' thickness $L_z$ and the large-scale velocity $U_{\rm L}$, using only the input parameters of the simulations. These predictions agree with our simulations, even outside the weakly nonlinear regime. It is also consistent with the idea that the flow selects $L_z$ and $U_{\rm L}$ such that the vertical Froude number $U_{\rm L} /(NL_z)$ is of order unity \citep{billant_self_2001}. For strongly stratified simulations at large Reynolds numbers, we observe that the measured wave frequency is impacted by a Doppler shift, commonly observed in stratified and rotating flows \citep{dileoni2015absorption,campagne2015disentangling,lam2020partitioning}.

2D stratified turbulence with shear modes, and 3D stratified turbulence are naturally more complex than the idealized simulations presented in this study because other instabilities can occur \citep{caulfield2021layering}. Yet, we expect our order of magnitude to be relevant for discussing some transitions of strongly stratified turbulence. In particular, the layering and the Doppler shift are absent from our simulation when the random forcing perturbs the layering because $\kc \lesssim \kfmax$ lies in the forcing range. This behavior, together with our estimate $\kc \propto Fr^{-3/5}/L$, may help to explain the emergence of layering in realistic flows and the discrepancies between simulations/experiments employing different setups. It is tempting to apply our argument to rotating flows without geostrophic modes. In that context, $\kc = Ro^{-3/5}/L$ where $Ro = U/(2 \Omega L)$ is the Rossby number and $\Omega$ the rotation rate. In that context, $L_h = 2\pi/\kc$ would correspond to the radius of nearly vertical columnar flows. 

For observing weak wave turbulence of internal waves, one must ensure that the large-scale flow, whose size is fixed by discrete wave turbulence \cite{nazarenko2011wave}, is destroyed. It can be done by adding large-scale damping \citep{lereun2017inertial,brunet2020}, or by using a random forcing over the all discrete turbulence range $k \lesssim \kc$. Another way of preventing the formation of large scale would be to use a large aspect ratio so the large scale flow cannot exist in the simulation box or the container. Interestingly, considering an asymptotic limit of the Navier-Stokes equations, \citet{vankan2020critical, vankan2022energy} identified transitions to large-scale flow formations for critical values of the aspect ratio. Yet, as explained by the authors, their asymptotic is different from weak wave turbulence and is unlikely to be directly related to the present work.


\backsection[Acknowledgements]{We thank Giorgio Krstulovic for providing the structure for the code used in this work together with endless advice; Oliver B\"uhler, Sergey Nazarenko, Jalal Shatah, Gregory Falkovich, Pierre Augier, Paul Billant, and Colm-cille Caulfield for fruitful discussions.}

\backsection[Funding]{This work was supported by the Simons Foundation and the Simons Collaboration on Wave Turbulence. The numerical study was made possible thanks to New York University's Greene computing cluster facility. MS acknowledges additional financial support from the Schmidt Futures Foundation.}

\backsection[Declaration of interests]{The authors report no conflict of interest.}


\backsection[Author ORCIDs]{
	V. Labarre, \href{https://orcid.org/0000-0002-5081-4008}{https://orcid.org/0000-0002-5081-4008}; M. Shavit, \href{https://orcid.org/0000-0002-7252-7077}{https://orcid.org/0000-0002-7252-7077}}.

\backsection[Author contributions]{Both authors conducted this study, performed the simulations, and wrote the manuscript.}



\bibliographystyle{jfm}
\bibliography{jfm}

\end{document}